# The Pseudo Radiation Energy Amplifier (PREA) and the mean earth's ground temperature

**Ahmed Boucenna**

Laboratoire DAC, Département de Physique
Faculté des Sciences, Université Ferhat Abbas, 19 000 Sétif, Algeria

………………………………………………………………...

**From the radiation balance diagram[1] illustrating the IPCC reports[2] one can estimate the power received by Earth from the sun at $P_{in}$ = 342 W/m$^2$ and the power consumed, remitted and reflected by the earth and its atmosphere at $P_{out}$ = 599 kW/m$^2$. It seems that the earth emits more power than it receives. The earth's ground mean temperature is estimated at 15 °C. A calculation based on the black body radiation theory gives an earth's ground mean temperature of the order of -18 °C which is much lower than 15 °C. The important gap between these calculated and estimated temperature mean values requires an explanation. Here we show that a gray body separated from vacuum by an interface and submitted to outside incident radiation can behave like a Pseudo Radiation Energy Amplifier. The Earth which is a gray body separated from the space by an interface, behaves like a Pseudo Radiation Energy Amplifier. The balance of the energy exchanged between Earth and outer space is reconsidered and the 15 °C Earth's ground temperature mean value is then derived. Our result revives the discussion on the parameters that control the Earth climate. The solar and terrestrial radiation reflection coefficients by the earth and its atmosphere acquire a privileged role. The Pseudo Radiation Energy Amplifier can be a starting point to improve the energy use. The materials having the adequate reflection coefficients will be indicated for specific applications.**

If a gray body receives a $P_0$ power radiation from an outside source and if R is the total mean reflection coefficient at the AB interface separating the gray body from the vacuum (Figure 1), then the $RP_0$ power is reflected and the power :

$$P = (1-R) P_0 \qquad (1)$$

is absorbed by the gray body.

The gray body radiates towards the vacuum the power $(1-R)P_0$ that it absorbed. This power tries to cross the AB interface characterized by a total mean reflection coefficient r for the infrared radiation (IR). Only a part $(1-r)(1-R)P_0$ crosses the AB interface, whereas the power $P_1 = r(1-R)P_0$ is reflected by the AB interface to the gray body itself which absorbs and reemits this power to the vacuum. The AB interface will then reflect to the gray body the power $P_2 = r\,P_1 = r^2(1-R)P_0$, and so on.

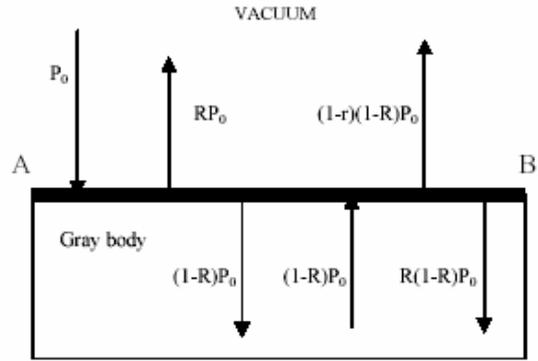

**Figure 1** The Pseudo Radiation Energy Amplifier (PREA). A gray body receives a $P_0$ power from an outside source. Radiations emitted by the different points of the heated gray body are reflected by the AB interface with a total mean reflection coefficient r. The AB interface can have variable reflection coefficients. In the case of the earth's ground, the AB interface is constituted by the Earth atmosphere.

At the end of the n$^{th}$ reflection, the power reflected by the AB interface is $P_n = r\,P_{n-1} = r^n(1-R)P_0$. Therefore, the gray body did not absorb solely the power $(1-R)P_0$ but it absorbed in addition a part of the power that is itself radiated and that is reflected back by the AB interface to the gray body. Then, the effective power absorbed by the gray body is:

$$\begin{aligned}P &= (1-R)P_0 + P_1 + P_2 + P_3 + \ldots + P_n + \ldots \\ &= (1-R)(P_0 + r^1 P_0 + r^2 P_0 + \ldots + r^n P_0 + \ldots)\end{aligned}$$

The power P effectively absorbed by the gray body and reemitted as IR radiations is equal to the sum of the geometric progression terms, where the first term is $(1-R)P_0$ and the reason is r :

$$P = \frac{1-R}{1-r} P_0 \qquad (2)$$

The power effectively absorbed by the gray body and reemitted as IR is not given by relation (1) but is given by relation (2). The total mean values of the R and r reflection coefficients depend on the wavelengths of incident radiation and the one reemitted by the gray body once it absorbed the power P. If the whole power absorbed is reemitted, the gray body must radiate towards the vacuum the power P that it absorbed. Depending on the mean values of the R and r reflection coefficients, the effectively P absorbed power by the gray body can be equal, lower or larger than the $P_0$ power received from the outside source. Thus:

- if $R = r$, then $P = P_0$, the effectively absorbed and remitted power by the gray body is equal to the power received from the outside source. The body behaves like a pseudo black body. The particular case of a black body corresponds to $R = r = 0$;

- if $R > r$, then $P < P_0$, the effectively absorbed and remitted power by the gray body is lower than the power received from the outside source, one has a loss of power ;

- if $R < r$, then $P > P_0$ the effectively absorbed and remitted power by the gray body is larger than the power received from the outside source, one has a gain of power. The body behaves like a Pseudo Radiation Energy Amplifier (PREA).

The absorbed power that gives the body its T temperature satisfies the Stefan Boltzmann law:

$$\sigma T^4 = \frac{1-R}{1-r} P_0 \qquad (3)$$

σ is the Stefan constant. If the temperature of a gray body is known one can deduce the radiated power and vice versa, if a gray body absorbs a P power one can deduce its T temperature.

The Pseudo Radiation Energy Amplifier is not an Energy Amplifier. The total energy reflected and remitted by the Pseudo Radiation Energy Amplifier, given by the relation:

$$\begin{aligned} P &= RP_0 + (1-r)(1-R)P_0 + (1-r)r(1-R)P_0 \\ &\quad + (1-r)r^2(1-R)P_0 + \ldots \\ &\quad + (1-r)r^n(1-R)P_0 + \ldots \\ &= RP_0 + (1-r)\frac{1-R}{1-r}P_0 \\ &= P_0 \end{aligned} \qquad (4)$$

is equal to the energy received from the outside source.

The power S received by the earth from the sun at the equator is given by:

$$S = \sigma T_{sun}^4 \frac{A_{sun}}{A_{earth\_orbit}} = \sigma T_{sun}^4 \frac{R_{sun}^2}{R_{earth\_orbit}^2} \qquad (5)$$

where $T_{sun}$ = 5780 °K. $R_{sun}$ and $R_{earth\_orbit}$ are respectively the sun and the earth orbit radii. $A_{sun}$ and $A_{earth\_orbit}$ are the sun and the earth orbit areas.

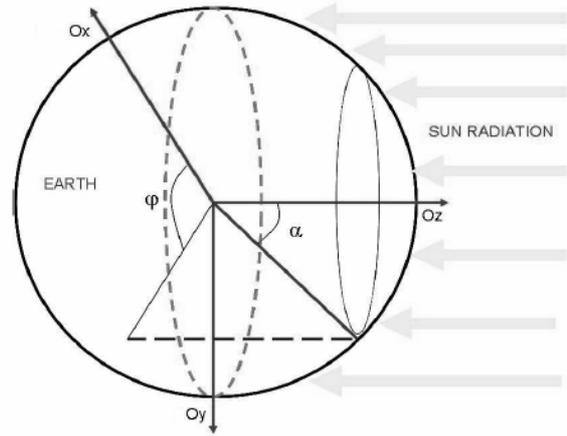

**Figure 2** Exposed Earth face. The α latitude is the angle between the Oz axis (passing by the Earth center and crossing the equator) and the vector locating the solar radiation impact point (on the Earth's ground).

At any α latitude point of the Earth exposed face, the power coming from the sun (Figure 2) is given by :

$$P = S \cos \alpha \qquad (6)$$

The mean power coming from the sun on the whole Earth's ground, $P_0$, is then:

$$\begin{aligned} P_0 &= \frac{1}{A} \int_{exposed\_face\_area} S \cos \alpha \, dA \\ &= \frac{S}{A} \int_0^{2\pi} \int_0^{\frac{\pi}{2}} \cos \alpha \sin \alpha \, d\alpha \, d\varphi = \frac{S}{4} \end{aligned} \qquad (7)$$

where A is the Earth area. For S = 1368 W/m², one has $P_0$ = 342 W/m², which is the power value $P_{in}$ received from the sun and given in the references[1,2]. Assuming that the interface [(Earth+Atmosphere) – space] presents the R and r total mean reflection coefficients to solar and terrestrial IR radiations respectively, the power effectively absorbed by the Earth is given by relation (2). This absorbed power is equal to the $P_{out}$ power consumed, reemitted and reflected by Earth and its atmosphere, therefore:

$$P_{out} = \frac{1-R}{1-r} P_0 \quad (8)$$

The power values mentioned in references[1,2] give us a value of the total mean reflection coefficient R. One has $R = \frac{107}{342} \approx 0.31$. And using the relation (8) one deduces the r reflection coefficient total mean value $r \approx 0.61$. One can see that for the Earth, $R < r$, the effectively absorbed power by the Earth and remitted is greater than the power received from the sun. We have a gain of power. The Earth behaves like a Pseudo Radiation Energy Amplifier. The power effectively absorbed by the Earth's ground and reemitted as IR radiations must satisfy the Stefan Boltzmann law :

$$\sigma T_{mean}^4 = \frac{1-R}{1-r} P_0 \quad (9)$$

In fact, a fraction $S_u$ of this power is consumed by Earth. The power used in the ocean water evaporation is estimated at 78 W/m² s and the one used to heat the air is 24 W/m² s. If one takes into account this power loss, relation (9) becomes:

$$T_{mean}^4 = \frac{1}{\sigma}\left[\frac{1-R}{1-r} P_0 - S_u\right] \quad (10)$$

For R = 0.31, r = 0.61, $P_0$ = 342 W/m² and $S_u$ = 102 W/m², one can obtain : $T_{mean} = 306.89$ K $= 33.89\,°C$. This mean temperature value is much higher than the 15 °C Earth's ground temperature mean value. To have this temperature, one must take r = 0.521. This reflection coefficient verifies the relation $R < r$ which means that the Earth behaves like a Pseudo Radiation Energy Amplifier. Using this value and relation (8), one can recalculate the $P_{out}$ power. One finds $P_{out} \approx 492$ W/m², instead of $P_{out} \approx 599$ W/m² given by the references[1,2]. This latter value is probably overestimated.

The power effectively absorbed by a Pseudo Radiation Energy Amplifier is the sum of the power provided by an external source and the power emitted by the gray body itself and reflected back on the interface separating the Pseudo Energy Amplifier from the vacuum. The solar power absorbed by the Earth does not explain the gap noted between the estimated Earth's ground temperature mean value and the one calculated while assimilating the earth to a black body. The Earth is in fact a Pseudo Radiation Energy Amplifier with adequate reflection coefficients. The Pseudo Radiation Energy Amplifier can find its application for a better use of the energy resources.

1. Kiehl, J. T. and Trenberth, K. E., 1997. Earth's Annual Global Mean Energy Budget, Bull. Amer. Meteor. Soc. , **78**, 197-208 (1997).

2. Intergovernmental Panel on Climate Change, IPCC, Rapports d'évaluation, (1990, 1995, 2001, 2007).

**Acknowledgements**

Thanks are due to Professor A. Layadi for his help.